\documentclass[english,showpacs,preprintnumbers,amsmath,amssymb]{revtex4}
\usepackage[latin9]{inputenc}
\usepackage{babel}
\usepackage{amsmath}
\usepackage{amssymb}
\usepackage{graphicx}
\usepackage[unicode=true,
 bookmarks=true,bookmarksnumbered=false,bookmarksopen=false,
 breaklinks=false,pdfborder={0 0 1},backref=false,colorlinks=false]
 {hyperref}

\makeatletter

\DeclareTextSymbolDefault{\textquotedbl}{T1}

\@ifundefined{textcolor}{}
{%
 \definecolor{BLACK}{gray}{0}
 \definecolor{WHITE}{gray}{1}
 \definecolor{RED}{rgb}{1,0,0}
 \definecolor{GREEN}{rgb}{0,1,0}
 \definecolor{BLUE}{rgb}{0,0,1}
 \definecolor{CYAN}{cmyk}{1,0,0,0}
 \definecolor{MAGENTA}{cmyk}{0,1,0,0}
 \definecolor{YELLOW}{cmyk}{0,0,1,0}
}

\usepackage{babel}
\usepackage{babel}

\makeatother

\begin{document}
\title{Mirrors and field sources in a Lorentz-violating scalar field theory}
\author{L. H. C. Borges}
\email{luizhenriqueunifei@yahoo.com.br}

\affiliation{UNESP - Campus de Guaratinguetá - DFQ, Avenida Dr. Ariberto Pereira
da Cunha 333, CEP 12516-410, Guaratinguetá, SP, Brazil}
\author{A. F. Ferrari}
\affiliation{Universidade Federal do ABC, Centro de Ciências Naturais e Humanas,
Rua Santa Adélia, 166, 09210-170, Santo André, SP, Brasil}
\email{alysson.ferrari@ufabc.edu.br}

\affiliation{Indiana University Center for Spacetime Symmetries, Indiana University,
Bloomington, Indiana 47405-7105, United States of America}
\author{F. A. Barone}
\email{fbarone@unifei.edu.br}

\affiliation{IFQ - Universidade Federal de Itajubá, Av. BPS 1303, Pinheirinho,
Caixa Postal 50, 37500-903, Itajubá, MG, Brazil}
\begin{abstract}
In this paper we consider classical effects in a model for a scalar
field incorporating Lorentz symmetry breaking due to the presence
of a single background vector $v^{\mu}$ coupled to its derivative.
We investigate of the interaction energy between stationary steady
sources concentrated along parallel branes with an arbitrary number
of dimensions, and derive from this study some physical consequences.
For the case of the scalar dipole we show the emergence of a nontrivial
torque, which is a distinctive sign of the Lorentz violation. We also
investigate a similar model in the presence of a semi-transparent
mirror. For a general relative orientation between the mirror and
the $v^{\mu}$, we are able to perform calculations perturbatively
in $v^{\mu}$ up to second order, and we also present exact results
specific cases. For all these configurations, the propagator for the
scalar field and the interaction force between the mirror and a point-like
field source are computed. It is shown that the image method is valid
in our model for the Dirichlet's boundary condition, and we argue
that this is a non-trivial result. We also show the emergence of a
torque on the mirror depending on its orientation with respect to
the Lorentz violating background: this is a new effect with no counterpart
in theories with Lorentz symmetry in the presence of mirrors.
\end{abstract}
\maketitle

\section{Introduction\label{I}}

\global\long\def\dimt{D+D_{\perp}+1}%
\global\long\def\dims{D}%
\global\long\def\dimp{D_{\perp}}%
Lorentz symmetry violating (LV) field theories received substantial
attention as a possible signature for underlying new physics arising
from the Planck scale. The search for Lorentz violation effects have
been developed in several branches of physics mainly in the framework
of the Standard Model Extension (SME) \cite{SME1,SME2,SME3,SME4}:
we mention, for instance, QED effects \cite{QED1,QED2,QED3,QED4,QED5,QED6,PetrQED1,PetrQED2,BJPMariz},
radiative corrections \cite{R1,R2,R3}, the study of Lorentz symmetry
violation with boundary conditions \cite{LHCFAB1}, effects in classical
electrodynamics \cite{cl1,cl2,cl3,cl4,BJPBorgesBarone,PRDHelayel,EPJCManoel},
Casimir effect \cite{CasimirFermion,Casimir1,Casimir2}, and effects
in the hydrogen atom \cite{BBhidrogenio}, among many others. Models
which exhibit Lorentz symmetry breaking and higher order derivatives
have also been studied\,\cite{BBF,Manoel1,Manoel2,ManoelARXIV,Petr1,Petr2,Petr3}.
In particular, scalar fields are especially interesting for exploring
the fundamental theoretical properties of field theories with Lorentz
violation\,\cite{scalar1,scalar2,scalar3,scalar4,scalar5,scalar6,scalar7,scalar8,scalar9,scalar10,scalar11,scalar12}
and, for the case of the Higgs fields, also for phenomenology\,\cite{higgs1,higgs2}.

Some recent works\,\cite{petrov1,petrov2} considered a model composed
by a massive real scalar field with an aether-like CPT-even Lorentz
symmetry breaking term, which is a coupling between the derivative
of the scalar field and a constant background vector $v^{\mu}$, and
studying the Casimir effect both for zero\,\cite{petrov1} and finite
temperature\,\cite{petrov2}. Inspired by these works, also using
a scalar field as the theoretical setup, one of the most fundamental
questions one can ask concerns the physical phenomena produced by
the presence of point-like sources, mainly the possible emergence
of phenomena with no counterpart in the Lorentz invariant case. A
related question concerns the modifications the Lorentz violating
scalar field propagator undergoes due to the presence of a single
semi-transparent-mirror, and its influence on static point-like field
sources. These questions deserve investigations not only for their
theoretical aspects, but also because of their possible relevance
in the search for Lorentz symmetry breaking.

In this work, starting from the model studied in\,\cite{petrov1,petrov2},
we consider stationary delta-like currents which are taken to be distributed
along parallel $D$-branes, and calculate exactly their interaction
energy, deriving from it some interesting particular cases. The same
analysis is performed for a distribution of scalar dipoles. Finally,
we investigate some consequences in our Lorentz violating model due
to the presence of a two dimensional semi-transparent mirror in a
$3+1$ dimensional spacetime. The calculations can be performed perturbatively
for a general orientation of the mirror and the background vector.
Exact results are also obtained for two special cases: when the LV
vector has only components parallel to the mirror, and when it has
a single component perpendicular to the mirror. For all these configurations,
we obtain the propagator for the scalar field and the interaction
force between the plate and a point-like field source. We also compare
the interaction forces with the ones obtained in the free theory (without
the mirror) and we verify that the image method is valid in all the
situations considered, for Dirichlet's boundary condition. This is
a nontrivial result since, even if LV in this model clearly preserves
the linearity of the equations of motion, the image method is also
dependent of the symmetries of the problem, which are modified by
the presence of the LV background. We show that a new effect arises
when a point-like source is placed in the vicinity of the mirror,
namely the existence of a small torque on the mirror, depending on
its position relative to the background vector. This is an effect
due to the Lorentz symmetry breaking, with no counterpart in standard
scalar field theory. Finally, we argue that, when we have the presence
of the mirror, the LV term cannot be eliminated with a coordinates
change.

The paper is organized as follows: in Section \ref{sec:Charges},
we develop a general setup considering effects of the presence of
$N$ stationary field sources (scalar charges and dipoles distributions)
concentrated at distinct regions of space, for arbitrary dimensions.
In Section \ref{sec:Mirror}, where we have the main results of the
paper, we compute, in a $3+1$ spacetime, the propagator for the scalar
field in the presence of a semi-transparent mirror considering different
configurations for the background vector. We use these results to
study the interaction energy between a point-like scalar charge and
the mirror in Section \ref{Interaction1}. We obtain some new results,
particularlya spontaneous torque acting on a setup where the distance
between the charge and the mirror is kept fixed. Section \ref{conclusions}
is dedicated to the conclusions and final remarks.

\section{Interaction between external sources\label{sec:Charges}}

In this section we shall deal with a model in $\dimt$ spacetime dimensions,
where $\dims$ will denote the dimensionality of the sources considered,
$\dimp$ will be the number of orthogonal space directions, and the
remaining coordinate $x^{0}$ represents time. It will be convenient
to denote by ${\bf x}_{\perp}$ and ${\bf x}_{\|}$ the space directions
perpendicular and parallel to the sources, so that the position four-vector
is given by $x^{\mu}=\left(x^{0},{\bf x}_{\perp},{\bf x}_{\|}\right)$.
We shall also use similar notations for the momenta $p^{\mu}$, as
well as for any other four-vector whenever necessary. The spacetime
metric is $\eta^{\mu\nu}=\text{diag}(+1,-1,-1,\ldots,-1)$. We shall
be dealing with sources represented by delta functions of different
dimensions (or derivatives of those), representing charges evenly
distributed on $\dims$ dimensional branes, in the most general sense.
Some particular cases will be considered after general results are
obtained. To avoid the problematic case of coinciding sources, we
shall always consider that $\dimp=1,2,3,\ldots$, while $D$ can be
any integer, including zero, which corresponds to point-like sources.

Let us consider a massive real scalar field $\phi$ in a Lorentz-symmetry
breaking scenario, defined by the following Lagrangian density\,\cite{petrov1,petrov2},
\begin{equation}
{\cal {L}}=\frac{1}{2}\partial_{\mu}\phi\partial^{\mu}\phi-\frac{1}{2}m^{2}\phi^{2}+\frac{1}{2}v^{\mu}v^{\nu}\partial_{\mu}\phi\partial_{\nu}\phi+J\phi\ ,\label{model1}
\end{equation}
where $m$ stands for the scalar field mass, $J$ is the external
source and $v^{\mu}$ is the Lorentz violating background vector which
is a dimensionless quantity, assumedly very small.

The scalar model considered by us can be related (in the massless
case) with the LV modification of electrodynamics studied in\,\cite{cl1}:
the two bosonic degrees of freedom of the electromagnetic field have
essentially the same dynamics as described by the massless limit of
Eq.\,\eqref{model1}, so we will be able to reproduce some of the
results presented in\,\cite{cl1}. The choice of the simplified scalar
modelwe consider allows to obtain more general, and even some exact,
results, at the price of not being directly comparable with experiments.
The Lorentz violating background is parametrized by a single vector
coefficient $v^{\mu}$, which justify the denomination of \textquotedbl aether-like\textquotedbl{}
scalar model used for example in\,\cite{petrov1,petrov2}. A general
parametrization for LV in a single scalar field theory have recently
been proposed in\,\cite{scalar12}, and the model studied by us can
be seen as a particular case of the minimal (involving only operators
of mass dimension not greater than four), CPT-even LV operator involving
the Klein-Gordon field denoted as 
\begin{equation}
{\cal L}_{LV}=\frac{1}{2}k_{c}^{\mu\nu}\partial_{\mu}\phi\partial_{\nu}\phi\ ,\label{genLVmin}
\end{equation}
where $k_{c}^{\mu\nu}$ can be considered to be traceless, since its
trace corresponds to a Lorentz invariant correction to the kinetic
term, which can be eliminated via a redefinition of the field and
the parameters of the theory. Our model corresponds to the particular
choice $k_{c}^{\mu\nu}=v^{\mu}v^{\nu}$. Notice that the tracelessness
condition of $k_{c}^{\mu\nu}$, in our particular case, is equivalent
to $v^{2}=v^{\mu}v_{\mu}=0$, which is a condition we can impose without
actually modifying any of the results we will present, except for
the calculation presented in the Appendix.

It is known that in a single-field theory, the LV contained in Eq.\,\eqref{genLVmin}
can actually be eliminated by means of a coordinate choice, absorbing
$k_{c}^{\mu\nu}$ in the spacetime metric itself\,\cite{Kos1,Kos2}.
However, in a general scenario, involving different fields and interactions
among them, this can be done for only one field at a time. Also, the
presence of the mirror, which by itself already breaks Lorentz invariance,
precludes the elimination of the LV by a coordinate redefinition.
This is why it is still important to investigate the consequences
of the LV described by Eq.\,\eqref{genLVmin}, since we can always
imagine the scalar field as belonging to a more complicated theory,
where we are actually not allowed, or it is not preferred to use this
freedom to eliminate $k_{c}^{\mu\nu}$ from the theory. We can still
use this freedom, however, to check the validity of one of our results,
as we will comment shortly.

The free propagator $G_{0}\left(x,y\right)$ is the inverse of the
kinetic operator ${\cal {O}}$, 
\begin{equation}
{\cal {O}}=\Box+m^{2}+\left(v\cdot\partial\right)^{2}\ ,\label{Operator1-1}
\end{equation}
which can be calculated by standard field theory methods. In the Fourier
representation, we can write 
\begin{equation}
G_{0}\left(x,y\right)=\int\frac{d^{\dimt}p}{\left(2\pi\right)^{\dimt}}\frac{e^{ip\cdot\left(x-y\right)}}{[p^{2}+\left(p\cdot v\right)^{2}-m^{2}]}\ .\label{propagator}
\end{equation}
This propagator is the basic ingredient we need to obtain several
physical quantities. For example, since the theory is quadratic in
the field variables $\phi$, it can be shown that the contribution
of the source $J\left(x\right)$ to the vacuum energy of the system
is given by\,\cite{fabarone2,fabarone3} 
\begin{equation}
E=\frac{1}{2T}\int\int d^{\dimt}x\ d^{\dimt}y\ J\left(x\right)G_{0}\left(x,y\right)J\left(y\right)\ ,\label{energy}
\end{equation}
where $T\rightarrow\infty$, $T$ being the time coordinate.

\subsection{Charges Distributions}

As discussed in \cite{fabarone2,fabarone3}, a stationary and uniform
scalar charge distribution lying along $D$-dimensional parallel branes
can be described by the external source 
\begin{equation}
J_{I}\left(x\right)=\sum_{k=1}^{N}\sigma_{k}\delta^{\dimp}\left({\bf {x}}_{\perp}-{\bf {a}}_{k}\right)\ ,\label{Current1}
\end{equation}
where ${\bf {a}}_{k}$, $k=1,\ldots,N$, are $N$ fixed $\dimp$-dimensional
spatial vectors describing the position of the branes in the transversal
space, and the parameters $\sigma_{k}$ are the coupling constants
between the field and the delta functions, playing the physical role
of generalized charge densities on the branes. Substituting (\ref{Current1})
into (\ref{energy}), discarding the self-interacting energies, we
have 
\begin{equation}
E_{I}=\frac{1}{2T}\sum_{k=1}^{N}\sum_{l=1}^{N}\sigma_{k}\sigma_{l}\left(1-\delta_{kl}\right)\int\int d^{\dimt}x\ d^{\dimt}y\ \delta^{\dimp}\left({\bf {x}}_{\perp}-{\bf {a}}_{k}\right)G_{0}\left(x,y\right)\delta^{\dimp}\left({\bf {y}}_{\perp}-{\bf {a}}_{l}\right)\ ,\label{energy2}
\end{equation}
where $\delta_{kl}$ is the Kronecker delta. This expression can be
simplified by using Eq. (\ref{propagator}) and computing the integrals
in the following order, $d^{\dimp}{\bf {x_{\perp}}}$, $d^{\dimp}{\bf {y_{\perp}}}$,
$dx^{0}$, $d^{\dims}{\bf {x_{\parallel}}}$, then introducing the
Fourier representation for the Dirac delta function and integrating
in $dp^{0},d^{\dims}{\bf {p_{\parallel}}}$, identifying the time
interval as $T=\int dy^{0}$, and $L^{\dims}=\int d^{\dims}{\bf {x_{\parallel}}}$
as being the volume of a given brane. After these manipulations, we
obtain 
\begin{equation}
{\cal {E}}_{I}=\frac{E_{I}}{L^{\dims}}=-\frac{1}{2}\sum_{k=1}^{N}\sum_{l=1}^{N}\sigma_{k}\sigma_{l}\left(1-\delta_{kl}\right)\int\frac{d^{\dimp}{\bf {p_{\perp}}}}{(2\pi)^{\dimp}}\frac{e^{-i{\bf {p}}_{\perp}\cdot{\bf {a}}_{kl}}}{[{\bf {p}}_{\perp}^{2}-\left({\bf {v}}_{\perp}\cdot{\bf {p}}_{\perp}\right)^{2}+m^{2}]}\ ,\label{energy3}
\end{equation}
where ${\bf {a}}_{kl}={\bf {a}}_{k}-{\bf {a}}_{l}$ and we have defined
${\cal {E}}_{I}$ as the energy per unit of $D$-brane volume.

In order to calculate the remaining integral in (\ref{energy3}),
we proceed as in\,\cite{LHCFAB1,scalar3} and take into account the
relative orientation of the vector ${\bf p}_{\perp}=\left(p^{1},\ldots,p^{\dimp}\right)$
and the spatial components perpendicular to the sources of the Lorentz
violating vector, i.e, ${\bf v}_{\perp}=\left(v^{1},\ldots,v^{\dimp}\right)$,
hence we split ${\bf p}_{\perp}$ into two parts, one parallel and
the other normal to ${\bf v}_{\perp}$, namely ${\bf p}_{\perp}={\bf p}_{\perp n}+{\bf p}_{\perp p}$,
where 
\begin{equation}
{\bf p}_{\perp p}={\bf v}_{\perp}\Bigl(\frac{{\bf v}_{\perp}\cdot{\bf p}_{\perp}}{{\bf v}_{\perp}^{2}}\Bigr),\ {\bf p}_{\perp n}={\bf p}_{\perp}-{\bf v}_{\perp}\Bigl(\frac{{\bf v}_{\perp}\cdot{\bf p}_{\perp}}{{\bf v}_{\perp}^{2}}\Bigr)\thinspace,\label{mudan1EM}
\end{equation}
so that ${\bf p}_{\perp n}\cdot{\bf v}_{\perp}=0$ by construction.
Now we define the vector ${\bf q}_{\perp}=\left(q^{1},\ldots,q^{\dimp}\right)$
as follows, 
\begin{equation}
{\bf q}_{\perp}={\bf p}_{\perp n}+{\bf p}_{\perp p}\sqrt{1-{\bf v}_{\perp}^{2}}\ .\label{defq}
\end{equation}
With these definitions one may write 
\begin{equation}
{\bf p}_{\perp p}=\frac{{\bf v}_{\perp}({\bf v}_{\perp}\cdot{\bf q}_{\perp})}{{\bf v}_{\perp}^{2}\sqrt{1-{\bf v}_{\perp}^{2}}}\ \ \ ,\ \ \ {\bf p}_{\perp n}={\bf q}_{\perp}-\frac{{\bf v}_{\perp}({\bf v}_{\perp}\cdot{\bf q}_{\perp})}{{\bf v}_{\perp}^{2}}\thinspace,\label{mudan6EM}
\end{equation}
leading to 
\begin{equation}
{\bf p}_{\perp}={\bf q}_{\perp}+\frac{({\bf {v}}_{\perp}\cdot{\bf q}_{\perp}){\bf {v}_{\perp}}}{{\bf v}_{\perp}^{2}}\left(\frac{1}{\sqrt{1-{\bf v}_{\perp}^{2}}}-1\right)\ ,
\end{equation}
and 
\begin{equation}
{\bf q}_{\perp}^{2}={\bf p}_{\perp}^{2}-({\bf v}_{\perp}\cdot{\bf p}_{\perp})^{2}\thinspace.\label{zxc2}
\end{equation}
Another definition which will be useful in what follows is 
\begin{equation}
{\bf b}_{kl}={\bf a}_{kl}+\left(\frac{1-\sqrt{1-{\bf v}_{\perp}^{2}}}{\sqrt{1-{\bf v}_{\perp}^{2}}}\right)\left(\frac{{\bf v}_{\perp}\cdot{\bf a}_{kl}}{{\bf v}_{\perp}^{2}}\right){\bf v}_{\perp}\ ,\label{zxc3}
\end{equation}
such that 
\begin{equation}
{\bf p}_{\perp}\cdot{\bf a}_{kl}={\bf b}_{kl}\cdot{\bf q}_{\perp}\thinspace.\label{mudan3EM}
\end{equation}
Finally, the Jacobian of the transformation from ${\bf p}$ to ${\bf q}$
can be calculated from (\ref{mudan6EM}), resulting in 
\begin{equation}
\det\left[\frac{\partial{\bf {p}_{\perp}}}{\partial{\bf {q}_{\perp}}}\right]=\frac{1}{\sqrt{1-{\bf v}_{\perp}^{2}}}\ .\label{mudan5EM}
\end{equation}

Putting all the previous expressions together, we end up with 
\begin{equation}
{\cal {E}}_{I}=-\frac{1}{2}\sum_{k=1}^{N}\sum_{l=1}^{N}\frac{\sigma_{k}\sigma_{l}\left(1-\delta_{kl}\right)}{\sqrt{1-{\bf v}_{\perp}^{2}}}\int\frac{d^{\dimp}{\bf {q_{\perp}}}}{(2\pi)^{\dimp}}\ \frac{e^{-i{\bf {q}}_{\perp}\cdot{\bf {b}}_{kl}}}{{\bf {q}}_{\perp}^{2}+m^{2}}\ ,\label{energy4}
\end{equation}
and now the integral can be solved exactly\,\cite{fabarone2}, leading
to 
\begin{equation}
{\cal {E}}_{I}=-\frac{1}{2}\frac{m^{\dimp-2}}{(2\pi)^{\dimp/2}}\frac{1}{\sqrt{1-{\bf v}_{\perp}^{2}}}\sum_{k=1}^{N}\sum_{l=1}^{N}\sigma_{k}\sigma_{l}\left(1-\delta_{kl}\right)\left(mb_{kl}\right)^{1-(\dimp/2)}K_{(\dimp/2)-1}\left(mb_{kl}\right)\ ,\label{energy5}
\end{equation}
where $K_{n}(x)$ stands for the K-Bessel function \cite{Arfken},
and 
\begin{equation}
b_{kl}=\mid{\bf {b}}_{kl}\mid=\sqrt{{\bf a}_{kl}^{2}+\frac{\left({\bf v}_{\perp}\cdot{\bf a}_{kl}\right)^{2}}{1-{\bf v}_{\perp}^{2}}}\ .\label{bkl}
\end{equation}

Expression (\ref{energy5}) is an exact result, which gives the interaction
energy per unit of D-brane volume between $N$ $D$-dimensional steady
and uniform field sources for the model. As expected, for $v^{\mu}=0$
or ${\bf v}_{\perp}=0$ expression (\ref{energy5}) reduces to the
standard Lorentz invariant result obtained in\,\cite{fabarone2}.
In the final result, the presence of the LV amounts to the dependence
of the energy not only on the perpendicular distance between the sources,
$a_{kl}$, but also on the orientation of the sources relative to
the LV vector ${\bf v}_{\perp}$.

It is interesting to notice that the possibility of removing the LV
from the theory via a coordinate choice allows us to find an alternative
derivation of this result, which serves as a consistency check. If
we consider the coordinate change 
\begin{equation}
x^{\mu}\rightarrow x^{\prime\mu}=x^{\mu}-\frac{1}{2}\left(v^{\nu}x_{\nu}\right)v^{\mu}\thinspace,\label{eq:coordchange}
\end{equation}
we can rewrite our model as a scalar theory living in a spacetime
with a modified metric given by, 
\begin{equation}
g^{\mu\nu}=\eta^{\mu\nu}-v^{\mu}v^{\nu}\thinspace,\label{eq:newmetric}
\end{equation}
in the first nontrivial order of $v$. Clearly, this metric effectively
absorbs the LV term present in Eq.\,\eqref{model1}, so our theory
is actually equivalent to the Lorentz invariant model given by 
\begin{equation}
S\left[\phi,J\right]=\int d^{4}x\thinspace\sqrt{-g}{\cal L}_{0}\left(\phi,\partial\phi,J\right)\thinspace,\label{eq:LIaction}
\end{equation}
where ${\cal L}_{0}$ corresponds to Eq.\,\eqref{model1} with $v=0$,
and we have dropped the primes on the new coordinates. The determinant
of the modified metric can be shown to be, in the first order, $\sqrt{-g}=\sqrt{1-v^{2}}$,
where we are not considering $v^{2}=0$ for reasons that will be clear
shortly. The determinant factor in Eq.\,\eqref{eq:LIaction} can
be absorbed by the rescaling 
\begin{equation}
\left(1-v^{2}\right)^{1/4}\phi\rightarrow\phi\thinspace,\thinspace\left(1-v^{2}\right)^{1/4}J\rightarrow J\thinspace.\label{eq:rescaling}
\end{equation}
The end result is a model identical to the one considered in\,\cite{fabarone2},
where LV have disappeared completely. The resulting energy can be
read from that reference, being given by 
\begin{equation}
{\cal {E}}_{LI}=-\frac{1}{2}\frac{m^{\dimp-2}}{(2\pi)^{\dimp/2}}\sum_{k=1}^{N}\sum_{l=1}^{N}\sigma_{k}\sigma_{l}\left(1-\delta_{kl}\right)\left(ma_{kl}\right)^{1-(\dimp/2)}K_{(\dimp/2)-1}\left(ma_{kl}\right)\ .\label{eq:EnergyLI}
\end{equation}

We can re-obtain (at the leading order) the result of the LV case,
Eq.\,\eqref{energy5}, by applying the inverse of the coordinate
choice\,\eqref{eq:coordchange}. One subtle point, however, is the
following: in deriving the energy density, we integrate over delta
functions of the form $\delta\left(p^{0}\right)$ and $\delta^{D}\left(\textbf{p}_{\parallel}\right)$,
which ends up eliminating all the dependency on the temporal and parallel
parts of $v^{\mu}$. As a result, in order to obtain our result, we
have to set $v^{\mu}\rightarrow\textbf{v}_{\perp}$. Therefore, we
consider the inverse coordinate choice as 
\begin{equation}
x_{\perp}^{i}\rightarrow x_{\perp}^{i\prime}=x_{\perp}^{i}+\frac{1}{2}v_{\perp}^{i}v_{\perp}^{j}x_{\perp}^{j}\thinspace,\label{eq:inverseCC}
\end{equation}
where the sum over repeated latin indices is implied. Applying this
transformation to the separation vector $\textbf{a}_{kl}$, we obtain
for the modulus of $\textbf{a}_{kl}^{\prime}$, 
\begin{equation}
a_{kl}^{\prime}=\sqrt{{\bf a}_{kl}^{2}+\left({\bf v}_{\perp}\cdot{\bf a}_{kl}\right)^{2}}\thinspace,
\end{equation}
where terms of higher order in $\textbf{v}_{\perp}$ were discarded.
This reproduces Eq.\,\eqref{bkl}, in the leading order. Finally,
the inverse of the rescaling\,\eqref{eq:rescaling} is 
\begin{equation}
\phi\rightarrow\left(1+v^{2}\right)^{-1/4}\phi\thinspace,\thinspace J\rightarrow\left(1+v^{2}\right)^{-1/4}J\thinspace,
\end{equation}
and, noticing that $\sqrt{1+v^{2}}\rightarrow\sqrt{1-{\bf v}_{\perp}^{2}}$,
we obtain the $\left(1-{\bf v}_{\perp}^{2}\right)^{-1/2}$ factor
present in Eq.\,\eqref{energy5}.

In order to gain insight into our results, we will now discuss some
particular cases. For the massless case, we have to consider separately
$\dimp=2$ and $\dimp\neq2$. Taking $m=0$ in (\ref{energy4}), the
relevant integral is written as 
\begin{equation}
{\cal {E}}_{I}\left(m=0\right)=-\frac{1}{2}\sum_{k=1}^{N}\sum_{l=1}^{N}\frac{\sigma_{k}\sigma_{l}\left(1-\delta_{kl}\right)}{\sqrt{1-{\bf v}_{\perp}^{2}}}\int\frac{d^{\dimp}{\bf {q_{\perp}}}}{(2\pi)^{\dimp}}\ \frac{e^{-i{\bf {q}}_{\perp}\cdot{\bf {b}}_{kl}}}{{\bf {q}}_{\perp}^{2}}\ ,\label{energy6}
\end{equation}
and for $\dimp\neq2$ we may directly integrate this expression, by
analytic continuation\,\cite{fabarone2}, obtaining 
\begin{align}
{\cal {E}}_{I}\left(m=0,\dimp\neq2\right)= & -\frac{2^{(\dimp/2)-3}}{(2\pi)^{\dimp/2}}\frac{1}{\sqrt{1-{\bf v}_{\perp}^{2}}}\ \Gamma\left(\frac{\dimp}{2}-1\right)\sum_{k=1}^{N}\sum_{l=1}^{N}\sigma_{k}\sigma_{l}\left(1-\delta_{kl}\right)\nonumber \\
 & \times\left[{\bf a}_{kl}^{2}+\frac{\left({\bf v}_{\perp}\cdot{\bf a}_{kl}\right)^{2}}{1-{\bf v}_{\perp}^{2}}\right]^{1-(\dimp/2)}\ ,\label{energy7}
\end{align}
with $\Gamma\left(x\right)$ standing for the Gamma Euler function.
For the specific case of $\dimp=2$, this last expression is divergent,
so a different regularization of the integral\,(\ref{energy6}) is
needed. We proceed as in\,\cite{cl1,fabarone2,fabarone3}, introducing
a mass regulator $M$, as follows 
\begin{equation}
{\cal {E}}_{I}\left(m=0,\dimp=2\right)=-\frac{1}{2}\sum_{k=1}^{N}\sum_{l=1}^{N}\frac{\sigma_{k}\sigma_{l}\left(1-\delta_{kl}\right)}{\sqrt{1-{\bf v}_{\perp}^{2}}}\lim_{M\rightarrow0}\int\frac{d^{2}{\bf {q_{\perp}}}}{(2\pi)^{2}}\ \frac{e^{-i{\bf {q}}_{\perp}\cdot{\bf {b}}_{kl}}}{{\bf {q}}_{\perp}^{2}+M^{2}}\ ,\label{energy8}
\end{equation}
so that we can use the integral \cite{fabarone2} 
\begin{equation}
\int\frac{d^{2}{\bf {q_{\perp}}}}{(2\pi)^{2}}\ \frac{e^{-i{\bf {q}}_{\perp}\cdot{\bf {b}}_{kl}}}{{\bf {q}}_{\perp}^{2}+M^{2}}=\frac{1}{2\pi}K_{0}\left(Mb_{kl}\right)\ ,\label{energy9}
\end{equation}
as well as the asymptotic expression of the Bessel function for small
arguments, 
\begin{align}
-K_{0}\left(Mb_{kl}\right) & =\ln\left(\frac{Mb_{kl}}{2}\right)+\gamma\ ,\label{bessel}\\
 & =\ln\left(\frac{b_{kl}}{a_{0}}\right)-\ln2+\gamma+\ln\left(Ma_{0}\right)\thinspace,
\end{align}
where $\gamma$ is the Euler constant and $a_{0}$ is an arbitrary
constant length scale. Terms that not depend on the distances $a_{kl}$
do not contribute to the force between the point-like currents, so
they can be discarded. We therefore arrive at 
\begin{equation}
{\cal {E}}_{I}\left(m=0,\dimp=2\right)=\frac{1}{4\pi\sqrt{1-{\bf v}_{\perp}^{2}}}\sum_{k=1}^{N}\sum_{l=1}^{N}\sigma_{k}\sigma_{l}\left(1-\delta_{kl}\right)\ln\left(\frac{b_{kl}}{a_{0}}\right)\ .\label{energy12}
\end{equation}
Notice that in these manipulations, we exchanged the dependence on
the arbitrary regulating mass $M$ for a regulating length $a_{0}$.
Despite explicitly appearing in Eq.\,\eqref{energy12} to keep the
argument of the logarithm dimensionless, $a_{0}$ does not appear
in derivatives of the energy, so it will not have any physical impacts.

In order to clarify the effects of the anisotropies generated by the
Lorentz-symmetry breaking, we will now consider some examples derived
from our general calculations. So, from now on we fix the dimensionality
of spacetime to be $3+1$, and the number of sources to be $N=2$.
When $\dimp=3,D=0$ we have two point-like sources in $3+1$ dimensions,
and the energy (\ref{energy5}) becomes 
\begin{equation}
{\cal {E}}_{I}\left(\dimp=3,D=0,N=2\right)=-\frac{\sigma_{1}\sigma_{2}}{4\pi\sqrt{1-{\bf v}^{2}}}\frac{e^{-mb}}{b}\ ,\label{energy13}
\end{equation}
where we discarded the sub-index $_{\perp}$ for simplicity, and 
\begin{equation}
b=b_{12}=b_{21}=\sqrt{{\bf a}_{12}^{2}+\frac{\left({\bf v}\cdot{\bf a}_{12}\right)^{2}}{1-{\bf v}^{2}}}=\sqrt{{\bf a}^{2}+\frac{\left({\bf v}\cdot{\bf a}\right)^{2}}{1-{\bf v}^{2}}}\ .\label{bkl12}
\end{equation}
If ${\bf v}=0$, the expression (\ref{energy13}) reduces to the well-known
Yukawa potential, otherwise the factor proportional to $\left({\bf v}\cdot{\bf a}\right)^{2}$
in the definition of $b$ in (\ref{bkl12}) implies in a dependence
of the energy on the relative orientation of the two charges and the
LV background. As a noteworthy particular case, if the distance vector
${\bf {a}}$ is perpendicular to the background vector ${\bf v}$,
Eq. (\ref{energy13}) becomes 
\begin{equation}
{\cal {E}}_{I}\left(\dimp=3,D=0,N=2,{\bf v}\cdot{\bf a}=0\right)=-\frac{\sigma_{1}\sigma_{2}}{4\pi\sqrt{1-{\bf v}^{2}}}\frac{e^{-m\mid{\bf {a}}\mid}}{\mid{\bf {a}}\mid}\ .\label{energy14}
\end{equation}
In this case the coefficient $1/\sqrt{1-{\bf v}^{2}}$ can be absorbed
into the definition of the coupling constants $\sigma_{1}$ and $\sigma_{2}$,
and Eq. (\ref{energy14}) reduces to the standard Yukawa potential.

Another interesting limit is the massless one, when we obtain 
\begin{equation}
{\cal {E}}_{I}\left(\dimp=3,D=0,N=2,m=0\right)=-\frac{\sigma_{1}\sigma_{2}}{4\pi\sqrt{1-{\bf v}^{2}}}\left[{\bf a}^{2}+\frac{({\bf v}\cdot{\bf a})^{2}}{1-{\bf v}^{2}}\right]^{-1/2}\ .\label{Ener6EM}
\end{equation}
This result can be directly compared with the one obtained in the
corresponding LV electrodynamics (EM) model studied in\,\cite{cl1}.
Equation\,(16) of\,\cite{cl1} presents the interaction energy between
two point-like charges in electrodynamics as
\begin{equation}
{\cal {E}}_{EM}=+\frac{\sigma_{1}\sigma_{2}}{4\pi}\frac{\sqrt{1-{\bf v}^{2}}}{1+v^{2}}\left[{\bf a}^{2}+\frac{({\bf v}\cdot{\bf a})^{2}}{1-{\bf v}^{2}}\right]^{-1/2}\thinspace.\label{eq:EMresult}
\end{equation}
Besides the expected minus sign relating the scalar and EM result,
one notices that the EM case depends on the temporal component $v^{0}$,
which decouples in the scalar model. Indeed, making $v^{0}=0$, the
result presented in Eq.\,\eqref{eq:EMresult} reproduces that of
Eq.\,\eqref{Ener6EM}, with a minus sign. In general, the same happens
for other quantities that we will calculate in the massless case,
enabling us to reobtain the results presented in\,\cite{cl1} as
particular cases of the calculations presented in this case.

The force between two point-like scalar charges can be calculated
from Eqs. (\ref{energy13}) and (\ref{bkl12}), resulting in 
\begin{align}
{\bf {F}}_{I}\left(\dimp=3,D=0,N=2\right) & =-{\bf {\nabla}}{\cal {E}}_{I}\left(\dimp=3,D=0,N=2\right)\nonumber \\
 & =-\frac{\sigma_{1}\sigma_{2}}{4\pi\sqrt{1-{\bf v}^{2}}}\frac{e^{-mb}}{b^{2}}\left(m+\frac{1}{b}\right)\left[{\bf {a}}+\frac{\left({\bf v}\cdot{\bf a}\right){\bf v}}{\sqrt{1-{\bf v}^{2}}}\right]\ ,\label{force1}
\end{align}
which depends on the direction of the background vector. When $m=0$,
the interaction force can be written in the following way 
\begin{equation}
{\bf {F}}_{I}\left(\dimp=3,D=0,N=2,m=0\right)=-\frac{\sigma_{1}\sigma_{2}}{4\pi{\bf a}^{2}}\frac{(1-{\bf v}^{2}){\hat{a}}+({\bf v}\cdot{\hat{a}}){\bf v}}{\left[1-{\bf v}^{2}+({\bf v}\cdot{\hat{a}})^{2}\right]^{3/2}}\ ,\label{for1}
\end{equation}
where ${\hat{a}}$ is an unit vector which points on the direction
of the distance vector ${\bf {a}}$.

Notice that (\ref{for1}) is an anisotropic force that decays with
the inverse square of the distance. In the special situation where
${\bf v}$ and ${\hat{a}}$ are perpendicular to each other, the force
(\ref{for1}) becomes a Coulombian-like interaction with effective
coupling constants $\sigma\to\sigma(1-{\bf v}^{2})^{-1/4}$. Since
$v$ is a small quantity, it is relevant to expand expression (\ref{for1})
in the lowest order in $v^{\mu}$, 
\begin{equation}
{\bf {F}}_{I}\left(\dimp=3,D=0,N=2,m=0\right)\cong-\frac{\sigma_{1}\sigma_{2}}{4\pi}\frac{1}{{\bf a}^{2}}\Biggl[\left(1+\frac{1}{2}{\bf v}^{2}-\frac{3}{2}\left({\bf v}\cdot{\hat{a}}\right)^{2}\right){\hat{a}}+({\bf v}\cdot{\hat{a}}){\bf v}\Biggr]\ .\label{for1exp}
\end{equation}
The first term inside the brackets is proportional to ${\hat{a}}$,
is a force in the same direction of the Lorentz invariant case, but
modulated by a function of the angle between ${\bf a}$ and ${\bf v}$,
the second term, however, is a new contribution proportional to the
LV vector ${\bf v}$ itself.

An interesting consequence of the anisotropy in the interaction energy
(\ref{energy13}) is the emergence of an spontaneous torque on a scalar
dipole, depending on its orientation relative to the LV background.
To see this, we consider a typical scalar dipole composed by two opposite
coupling constants $\sigma_{1}=-\sigma_{2}=\sigma$, placed at positions
${\bf {a}}_{1}={\bf {R}}+\frac{{\bf {A}}}{2}$ and ${\bf {a}}_{2}={\bf {R}}-\frac{{\bf {A}}}{2}$,
${\bf {A}}$ taken to be a fixed vector. From Eq.\,(\ref{energy13}),
we obtain 
\begin{equation}
{\cal {E}}_{I}^{dipole}\left(\dimp=3,D=0,N=2\right)=\frac{\sigma^{2}}{4\pi\sqrt{1-{\bf v}^{2}}}\frac{e^{-m\mid{\bf {A}}\mid f(\theta)}}{\mid{\bf {A}}\mid f(\theta)}\ ,\label{torque1}
\end{equation}
where 
\begin{equation}
f(\theta)=\sqrt{1+\frac{{\bf v}^{2}\cos^{2}(\theta)}{1-{\bf v}^{2}}}\ ,\label{ftheta}
\end{equation}
and $0\leq\theta\leq\pi$ stands for the angle between ${\bf {A}}$
and the background vector ${\bf {v}}$. This interaction energy leads
to an spontaneous torque on the dipole as follows, 
\begin{align}
\tau_{I}^{dipole} & \left(\dimp=3,D=0,N=2\right)=-\frac{\partial{\cal {E}}_{I}^{dipole}\left(\dimp=3,D=0,N=2\right)}{\partial\theta}\nonumber \\
 & =-\frac{\sigma^{2}}{8\pi\mid{\bf {A}}\mid}\frac{{\bf v}^{2}}{\left(1-{\bf v}^{2}\right)^{3/2}}\frac{1}{f^{2}(\theta)}\left(m\mid{\bf {A}}\mid+\frac{1}{f(\theta)}\right)\sin(2\theta)e^{-m\mid{\bf {A}}\mid f(\theta)}\ .
\end{align}

This spontaneous torque on the scalar dipole is an exclusive effect
due to the Lorentz violating background. If $v^{\mu}=0$ (or, more
specifically, ${\bf v}_{\perp}=0$), the torque vanishes, as it should,
as well as for the specific configurations $\theta=0,\pi/2,\pi$.
For the massless case the torque becomes 
\begin{align}
\tau_{I}^{dipole}\left(\dimp=3,D=0,N=2,m=0\right) & \cong-\frac{\sigma^{2}{\bf v}^{2}}{8\pi\mid{\bf {A}}\mid}\sin(2\theta)\ ,
\end{align}
which exhibits a maximum value at $\theta=\pi/4$. A similar effect
was also described in\,\cite{cl1,borges1}.

One interesting question regards possible phenomenological implications
of the presence of this LV induced torque on a dipole. Clearly, the
scalar model cannot be directly applied to any low energy experiments,
but as we mentioned, the results for the more realistic EM case are
very similar, and indeed this spontaneous torque of order $\sim\frac{\sigma^{2}{\bf v}^{2}}{8\pi\mid{\bf {A}}\mid}$
was also found in the EM calculation presented in\,\cite{cl1}. The
most obvious candidate for an experiment measuring such kind of torque
would be some kind of torsion pendulum, where sensibilities for torques
of order $10^{-16}\text{Nm}$ (or $10^{-26}\times1/\ell_{P}$ in natural
units, $\ell_{P}$ being the Planck length) are possible\,\cite{Shao:2015gua}.
However, this is still far from the order of magnitude of these induced
torques, which for a dipole of centimeter size, and with charge of
$n_{e}$ times the electron charge, would be of order
\begin{equation}
\tau_{LV}\sim{\bf v}^{2}n_{e}^{2}\times10^{-37}\times1/\ell_{P}\thinspace.
\end{equation}
Since ${\bf v}^{2}$ should be certainly many order of magnitude smaller
than unity, it is hard to imagine that $\tau_{LV}$ could be measured
with current technology.

The final examples we consider involve one and two dimensional sources,
i.e., strings and planes. For $\dimp=2,D=1$ and $N=2$ we have two
delta-like scalar charges distributions concentrated along two different
parallel strings placed at a distance ${\bf {a}}$ from each other.
In this case, from Eq.\,(\ref{energy5}) the energy per string length
reads 
\begin{equation}
{\cal {E}}_{I}\left(\dimp=2,D=1,N=2\right)=-\frac{\sigma_{1}\sigma_{2}}{2\pi\sqrt{1-{\bf v}_{\perp}^{2}}}K_{0}\left(mb\right)\ ,\label{enerd1}
\end{equation}
which is reduced, in the case $m=0$, to 
\begin{equation}
{\cal {E}}_{I}\left(\dimp=2,D=1,N=2,m=0\right)=-\frac{\sigma_{1}\sigma_{2}}{2\pi\sqrt{1-{\bf v}_{\perp}^{2}}}\ln\left(\frac{b}{a_{0}}\right)\ ,\label{enerd1m}
\end{equation}
where we used (\ref{energy12}).

Finally, for $\dimp=1,D=2$ and $N=2$, corresponding to two delta
currents concentrated on parallel planes, we have 
\begin{equation}
{\cal {E}}_{I}\left(\dimp=1,D=2,N=2\right)=-\frac{\sigma_{1}\sigma_{2}}{2m\sqrt{1-{\bf v}_{\perp}^{2}}}e^{-mb}\ ,\label{enerd2}
\end{equation}
or, in the massless limit, 
\begin{equation}
{\cal {E}}_{I}\left(\dimp=1,D=2,N=2,m=0\right)=\frac{\sigma_{1}\sigma_{2}}{2\sqrt{1-{\bf v}_{\perp}^{2}}}\sqrt{{\bf a}^{2}+\frac{\left({\bf v}_{\perp}\cdot{\bf a}\right)^{2}}{1-{\bf v}_{\perp}^{2}}}\ .\label{enerd2m}
\end{equation}

\subsection{Point-like Dipoles\label{sec:Dipoles}}

The technique developed in this section can be applied to other interesting
systems, such as dipole distributions, when the relevant currents
involve derivatives of delta functions. In this subsection we provide
some results in the case of two steady point-like dipoles placed at
fixed points in $3+1$ dimensions. This setup can be described by
external sources given by the directional derivatives of the Dirac
delta function\,\cite{fabarone2}, as follows 
\begin{equation}
J_{II}\left(x\right)=\sum_{k=1}^{2}\ {\bf V}_{(k)}\cdot{\nabla}\left[\delta^{3}\left({\bf {x}}-{\bf {a}}_{k}\right)\right]\ ,\label{Current2}
\end{equation}
where ${\bf V}_{(k)}^{\mu}$ designates the dipole moments $1$ and
$2$, taken to be fixed in the reference frame in which we are performing
the calculations. Following the same steps presented in the previous
section, we obtain for the interaction energy between the two dipoles,
\begin{equation}
E_{II}=-\int\frac{d^{3}{\bf {p_{\perp}}}}{(2\pi)^{3}}\ e^{-i{\bf {p}}_{\perp}\cdot{\bf {a}}}\frac{\left({\bf {V}}_{(1)\perp}\cdot{\bf {p}}_{\perp}\right)\left({\bf {V}}_{(2)\perp}\cdot{\bf {p}}_{\perp}\right)}{[{\bf {p}}_{\perp}^{2}-\left({\bf {v}}\cdot{\bf {p}}_{\perp}\right)^{2}+m^{2}]}\ ,\label{energydp1}
\end{equation}
where ${\bf a}={\bf a}_{1}-{\bf a}_{2}={\bf a}_{12}$.

Performing the same change in the integration variables as used in
the previous section, and adopted in\,\cite{LHCFAB1,scalar3}, using
the definition (\ref{zxc3}), we end up with 
\begin{multline}
E_{II}=\frac{1}{4\pi\sqrt{1-{\bf v}_{\perp}^{2}}}\frac{e^{-mb}}{b^{3}}\Biggl\{\frac{\left[\left(mb\right)^{2}+3\left(mb+1\right)\right]}{b^{2}}\Biggl[\left({\bf {V}}_{(1)\perp}\cdot{\bf {a}}\right)\left({\bf {V}}_{(2)\perp}\cdot{\bf {a}}\right)\\
+\frac{\left({\bf {v}}\cdot{\bf {a}}\right)}{1-{\bf v}^{2}}\left[\left({\bf {V}}_{(1)\perp}\cdot{\bf {a}}\right)\left({\bf {V}}_{(2)\perp}\cdot{\bf {v}}\right)+\left({\bf {V}}_{(2)\perp}\cdot{\bf {a}}\right)\left({\bf {V}}_{(1)\perp}\cdot{\bf {v}}\right)\right]\\
+\left(\frac{{\bf {v}}\cdot{\bf {a}}}{1-{\bf v}^{2}}\right)^{2}\left({\bf {V}}_{(1)\perp}\cdot{\bf {v}}\right)\left({\bf {V}}_{(2)\perp}\cdot{\bf {v}}\right)\Biggr]\\
-\left(mb+1\right)\Biggl[\left({\bf {V}}_{(1)\perp}\cdot{\bf {V}}_{(2)\perp}\right)+\frac{\left({\bf {V}}_{(1)\perp}\cdot{\bf {v}}\right)\left({\bf {V}}_{(2)\perp}\cdot{\bf {v}}\right)}{{1-{\bf v}^{2}}}\Biggr]\Biggr\}\ .
\end{multline}

In the massless case, we can use (\ref{zxc3}) and write 
\begin{multline}
E_{II}(m=0)=\frac{1}{4\pi\sqrt{1-{\bf v}^{2}}}\left[{\bf a}^{2}+\frac{\left({\bf v}\cdot{\bf a}\right)^{2}}{1-{\bf v}^{2}}\right]^{-3/2}\Biggl\{3\left[{\bf a}^{2}+\frac{\left({\bf v}\cdot{\bf a}\right)^{2}}{1-{\bf v}^{2}}\right]^{-1}\\
\times\Biggl[\left({\bf {V}}_{(1)\perp}\cdot{\bf {a}}\right)\left({\bf {V}}_{(2)\perp}\cdot{\bf {a}}\right)+\frac{\left({\bf {v}}\cdot{\bf {a}}\right)}{1-{\bf v}^{2}}\Bigl[\left({\bf {V}}_{(1)\perp}\cdot{\bf {a}}\right)\left({\bf {V}}_{(2)\perp}\cdot{\bf {v}}\right)+\left({\bf {V}}_{(2)\perp}\cdot{\bf {a}}\right)\left({\bf {V}}_{(1)\perp}\cdot{\bf {v}}\right)\Bigr]\\
+\left(\frac{{\bf {v}}\cdot{\bf {a}}}{1-{\bf v}^{2}}\right)^{2}\left({\bf {V}}_{(1)\perp}\cdot{\bf {v}}\right)\left({\bf {V}}_{(2)\perp}\cdot{\bf {v}}\right)\Biggr]-\left({\bf {V}}_{(1)\perp}\cdot{\bf {V}}_{(2)\perp}\right)-\frac{\left({\bf {V}}_{(1)\perp}\cdot{\bf {v}}\right)\left({\bf {V}}_{(2)\perp}\cdot{\bf {v}}\right)}{{1-{\bf v}^{2}}}\Biggr\}\thinspace.\label{energydp8}
\end{multline}
For the case where $v=0$ or ${\bf {v}}=0$, we have the well-known
result obtained in standard scalar field theory\,\cite{fabarone2},
\begin{equation}
E_{II}\left(m=v=0\right)=\frac{\sigma_{1}\sigma_{2}}{4\pi\mid{\bf {a}}\mid^{3}}\left[3\frac{\left({\bf {V}}_{(1)\perp}\cdot{\bf {a}}\right)\left({\bf {V}}_{(2)\perp}\cdot{\bf {a}}\right)}{{\bf {a}}^{2}}-\left({\bf {V}}_{(1)\perp}\cdot{\bf {V}}_{(2)\perp}\right)\right]\thinspace.\label{energydp9}
\end{equation}
Different particular cases can be analyzed, and torques depending
on the orientation of the dipoles relative to the LV background can
be deduced. Since these results follow directly from the approach
outlined in the previous subsection, we will not quote the explicit
expressions here.

\section{The propagator in the presence of a semi-transparent mirror\label{sec:Mirror}}

In this section we compute the propagator for the model (\ref{model1})
in the presence of a two-dimensional semi-transparent mirror. We keep
spacetime $3+1$ dimensional hereafter, and take a coordinate system
where the mirror is perpendicular to the $x^{3}$ axis, located on
the plane $x^{3}=0$. This configuration is described by the potential
$\frac{\mu}{2}\delta(x^{3})$, where $\mu>0$ is a coupling constant
with appropriate dimension, establishing the degree of transparency
of the mirror: the limit $\mu\rightarrow\infty$ corresponds to a
perfect mirror\,\cite{fabarone4,cavalcanti}. Therefore, the Lagrangian
density is given by 
\begin{equation}
{\cal {L}}=\frac{1}{2}\partial_{\mu}\phi\partial^{\mu}\phi-\frac{1}{2}m^{2}\phi^{2}+\frac{1}{2}\left(v\cdot\partial\phi\right)^{2}-\frac{1}{2}\mu\delta(x^{3})\phi^{2}+J\phi\ .\label{modelM}
\end{equation}

Here some comments are in order. The external delta-like potential
in the Lagrangian (\ref{modelM}) can be interpreted as a semi-transparent
mirror for the scalar field due to the following reasons: we can show
that the limit $\mu\to\infty$ of this coupling is equivalent to imposing
Dirichlet boundary conditions on the scalar field on the $x^{3}=0$
plane; besides, there is a close connection between the scalar field
with Dirichlet boundary conditions and the electromagnetic field in
the presence of a conducting plate, where the name mirror is more
appropriate. In fact, a model for a semi-transparent mirror with delta-like
potentials can also be established for the electromagnetic field\,\cite{BBplacacarga,BBCasimir}.
Finally, the presence of the delta function potential precludes the
elimination of the LV by means of a coordinate choice, since while
the $\left(v\cdot\partial\phi\right)^{2}$ term can be absorbed by
the kinetic term with the redefinition of the metric, the LV parameter
$v^{\mu}$ will reappear in the argument of the delta function potential
that represents the mirror. Actually, the $x^{3}=0$ plane in the
original coordinates will be in general mapped to a new plane in $3+1$
spacetime, with $v$ dependent orientation.

The propagator $G(x,y)$ for this theory satisfies the differential
equation 
\begin{equation}
\left[\Box+m^{2}+\left(v\cdot\partial\right)^{2}+\mu\delta(x^{3})\right]G(x,y)=-\delta^{4}(x-y)\ ,\label{Operator1M}
\end{equation}
and also a kind of Bethe-Salpeter equation 
\begin{equation}
G(x,y)=G_{0}(x,y)+\int d^{4}z\ G(x,z)\mu\delta(z^{3})G_{0}(z,y)\ ,\label{G1}
\end{equation}
where $G_{0}(x,y)$ is the free propagator given by the Eq. (\ref{propagator}),
which solves (\ref{Operator1M}) without the potential. From now on,
we define $x_{p}^{\mu}=(x^{0},x^{1},x^{2})$ and $p_{p}^{\mu}=(p^{0},p^{1},p^{2})$
as the coordinates and momentum parallel to the mirror, respectively.

In order to solve Eq. (\ref{Operator1M}), it is convenient to write
$G(x,y)$ and $G_{0}(x,y)$ as Fourier transforms in the parallel
coordinates, as follows, \begin{subequations} \label{G2} 
\begin{align}
G(x,y) & =\int\frac{d^{3}p_{p}}{(2\pi)^{3}}e^{ip_{p}\cdot(x_{p}-y_{p})}{\cal {G}}(p_{p};x^{3},y^{3})\ ,\\
G_{0}(x,y) & =\int\frac{d^{3}p_{p}}{(2\pi)^{3}}e^{ip_{p}\cdot(x_{p}-y_{p})}{\cal {G}}_{0}(p_{p};x^{3},y^{3})\ ,
\end{align}
\end{subequations} where ${\cal {G}}(p_{p};x^{3},y^{3})$ and ${\cal {G}}_{0}(p_{p};x^{3},y^{3})$
stand for the reduced Green's functions\,\cite{fabarone4,cavalcanti}.
Substituting (\ref{G2}) in (\ref{G1}) and performing some manipulations
we arrive at 
\begin{equation}
{\cal {G}}(p_{p};x^{3},y^{3})={\cal {G}}_{0}(p_{p};x^{3},y^{3})+\mu{\cal {G}}(p_{p};x^{3},0){\cal {G}}_{0}(p_{p};0,y^{3})\thinspace.\label{G3}
\end{equation}

Setting $y^{3}=0$ in (\ref{G3}), we can obtain ${\cal {G}}(p_{p};x^{3},0)$
strictly in terms of ${\cal {G}}_{0}(p_{p};x^{3},0)$. Using the result
back again in Eq. (\ref{G3}), we obtain 
\begin{equation}
{\cal {G}}(p_{p};x^{3},y^{3})={\cal {G}}_{0}(p_{p};x^{3},y^{3})+\frac{\mu{\cal {G}}_{0}(p_{p};x^{3},0){\cal {G}}_{0}(p_{p};0,y^{3})}{1-\mu{\cal {G}}_{0}(p_{p};0,0)}\ ,\label{G4}
\end{equation}
Substituting (\ref{G4}) in equation (\ref{G2}) leads to 
\begin{equation}
G(x,y)=G_{0}(x,y)+{\bar{G}}(x,y)\ ,\label{G5}
\end{equation}
where 
\begin{equation}
{\bar{G}}(x,y)=\int\frac{d^{3}p_{p}}{(2\pi)^{3}}e^{ip_{p}\cdot(x_{p}-y_{p})}\frac{\mu{\cal {G}}_{0}(p_{p};x^{3},0){\cal {G}}_{0}(p_{p};0,y^{3})}{1-\mu{\cal {G}}_{0}(p_{p};0,0)}\ .\label{G6}
\end{equation}

The propagator (\ref{G5}) is composed of the sum of the free propagator
(\ref{propagator}) with the correction (\ref{G6}), which accounts
for the presence of the semi-transparent mirror. Taking the limit
$\mu\to\infty$ in (\ref{G4}) and evaluating the resulting expression
for $x^{3}=0$, we can show that 
\begin{equation}
\lim_{\mu\to\infty}{\cal {G}}(p_{p};x^{3}=0,y^{3})={\cal {G}}_{0}(p_{p};0,y^{3})-\frac{{\cal {G}}_{0}(p_{p};0,0){\cal {G}}_{0}(p_{p};0,y^{3})}{{\cal {G}}_{0}(p_{p};0,0)}=0\ ,
\end{equation}
so the Green's function of the model satisfies the Dirichlet boundary
condition on the plane $x^{3}=0$ in the limit $\mu\to\infty$.In
this sense, we can interpret the delta-like external potential in\,(\ref{modelM})
as a kind of mirror, with degree of transparency given by $\mu$.

From now on, we will calculate ${\bar{G}}(x,y)$ for different configurations
of the mirror with respect to the background vector.

\subsection{The propagator in the lowest order in $v$\label{IV}}

Since $v^{\mu}$ is assumedly a very small parameter, we will perform
the calculations perturbatively up to the second order in $v^{\mu}$,
which is the lowest order in which the background vector appears non-trivially.
Expanding the propagator (\ref{propagator}), we obtain 
\begin{equation}
G_{0}(x,y)=\int\frac{d^{4}p}{(2\pi)^{4}}\frac{e^{ip\cdot(x-y)}}{(p^{2}-m^{2})}\left[1-\frac{\left(p\cdot v\right)^{2}}{(p^{2}-m^{2})}\right]\ .\label{G7}
\end{equation}
Splitting $G_{0}(x,y)$ into parallel and perpendicular coordinates
with respect to the mirror, we have 
\begin{equation}
G_{0}(x,y)=\int\frac{d^{3}p_{p}}{(2\pi)^{3}}e^{ip_{p}\cdot(x_{p}-y_{p})}\left[\int\frac{dp^{3}}{2\pi}\frac{e^{-ip^{3}(x^{3}-y^{3})}}{(p^{2}-m^{2})}\left(1-\frac{\left(p\cdot v\right)^{2}}{(p^{2}-m^{2})}\right)\right]\ ,\label{G8}
\end{equation}
where $p^{3}$ stands for the momentum perpendicular to the mirror.
From Eq. (\ref{G2}), we identify the term between brackets on the
right hand side of Eq. (\ref{G8}) as being ${\cal {G}}_{0}(p_{p};x^{3},y^{3})$.

The fact that \cite{fabarone4} 
\begin{eqnarray}
\int\frac{dp^{3}}{2\pi}\frac{e^{-ip^{3}(x^{3}-y^{3})}}{(p^{2}-m^{2})}=-\frac{e^{-\lambda\mid x^{3}-y^{3}\mid}}{2\lambda}\ ,\label{G9}
\end{eqnarray}
where $\lambda=\sqrt{m^{2}-p_{p}^{2}}$, leads to, 
\begin{align}
{\cal {G}}_{0}(p_{p};x^{3},y^{3}) & =-\frac{e^{-\lambda\mid x^{3}-y^{3}\mid}}{2\lambda}\Biggl\{1+\frac{1}{2}\Biggl[\frac{\left(p_{p}\cdot v_{p}\right)^{2}}{\lambda^{2}}\left(1+\lambda\mid x^{3}-y^{3}\mid\right)\nonumber \\
 & -2iv^{3}\left(x^{3}-y^{3}\right)\left(p_{p}\cdot v_{p}\right)+(v^{3})^{2}\left(1-\lambda\mid x^{3}-y^{3}\mid\right)\Biggr]\Biggr\}\ ,
\end{align}
with $v_{p}^{\mu}=(v^{0},v^{1},v^{2})$ and $v^{3}$ standing for
the background vector parallel and perpendicular to the mirror, respectively.
Substitution of this last expression into Eq. (\ref{G6}), and taking
into account contributions up to second order in $v^{\mu}$, provides
\begin{align}
{\bar{G}}(x,y) & =\int\frac{d^{3}p_{p}}{(2\pi)^{3}}e^{ip_{p}\cdot(x_{p}-y_{p})}\Biggl\{1+\frac{(p_{p}\cdot v_{p})^{2}}{2\lambda^{2}}\left[\left(\frac{4\lambda+\mu}{2\lambda+\mu}\right)+\lambda(\mid x^{3}\mid+\mid y^{3}\mid)\right]\nonumber \\
 & -iv^{3}(x^{3}-y^{3})(p_{p}\cdot v_{p})+\frac{(v^{3})^{2}}{2}\left[\left(\frac{4\lambda+\mu}{2\lambda+\mu}\right)-\lambda(\mid x^{3}\mid+\mid y^{3}\mid)\right]\Biggr\}\frac{\mu e^{-\lambda(\mid x^{3}\mid+\mid y^{3}\mid)}}{2\lambda(2\lambda+\mu)}\ .\label{G11}
\end{align}
As expected in this perturbative result, the limit $v^{\mu}\rightarrow0$
correctly reproduces the standard result for the scalar field theory
in the presence of a semi-transparent mirror\,\cite{fabarone4}.

\subsection{Exact propagators\label{V}}

There are two special cases for which we carry out the calculations
without the necessity of treating the background vector perturbatively,
corresponding to the spacial part of $v^{\mu}$ being parallel and
perpendicular to the mirror. In this subsection we present the exact
propagator in the presence of a semi-transparent mirror in these cases.

When the component of the background vector perpendicular to the mirror
is equal to zero ($v^{3}=0$) , we have (see the Appendix) 
\begin{equation}
{\cal {G}}_{0}(p_{p};x^{3},y^{3})=-\frac{e^{-L\mid x^{3}-y^{3}\mid}}{2L}\ ,\label{exactG1}
\end{equation}
where $L=\sqrt{m^{2}-\left[p_{p}^{2}+(p_{p}\cdot v_{p})^{2}\right]}$.
Substituting (\ref{exactG1}) in (\ref{G6}), we arrive at 
\begin{equation}
{\bar{G}}(x,y)=\int\frac{d^{3}p_{p}}{(2\pi)^{3}}e^{ip_{p}\cdot(x_{p}-y_{p})}\frac{\mu e^{-L(\mid x^{3}\mid+\mid y^{3}\mid)}}{2L(2L+\mu)}\thinspace.\label{exactG2}
\end{equation}

On the other hand, when $v_{p}^{\mu}=0$ and $v^{3}\neq0$, we can
write (see the Appendix) 
\begin{equation}
{\cal {G}}_{0}(p_{p};x^{3},y^{3})=-\frac{e^{-\lambda\left(\sqrt{1-(v^{3})^{2}}\right)^{-1}\mid x^{3}-y^{3}\mid}}{2\lambda\sqrt{1-(v^{3})^{2}}}\ ,\label{exactG1v3}
\end{equation}
what leads to 
\begin{equation}
{\bar{G}}(x,y)=\int\frac{d^{3}p_{p}}{(2\pi)^{3}}e^{ip_{p}\cdot(x_{p}-y_{p})}\frac{\mu e^{-\lambda\left(\sqrt{1-(v^{3})^{2}}\right)^{-1}(\mid x^{3}\mid+\mid y^{3}\mid)}}{2\lambda\sqrt{1-(v^{3})^{2}}\left(2\lambda\sqrt{1-(v^{3})^{2}}+\mu\right)}\ .\label{exactG2v3}
\end{equation}
It is easy to see that these expressions reproduce the result previously
obtained when expanded up to the second order in $v^{\mu}$.

\section{Charge-mirror interaction\label{Interaction1}}

Having calculated the relevant propagator in the previous section,
here we consider the interaction energy between a point-like current
and the semi-transparent mirror, which is given by\,\cite{fabarone4}
\begin{equation}
E=\frac{1}{2T}\int\int d^{4}x\ d^{4}y\ J(x){\bar{G}}(x,y)J(y)\ .\label{intmc1}
\end{equation}
Without loss of generality (due to translation invariance in the directions
parallel to the mirror) and for simplicity, we choose a point-like
scalar charge placed at ${\bf {a}}=(0,0,a)$, corresponding to the
source $J(x)=\sigma\delta^{3}({\bf {x}}-{\bf {a}})$. Again, we will
present a result perturbative in $v$ for the general case, and also
exact results for particular cases.

\subsection{Perturbative results\label{Interaction2}}

Expanding the expressions up to second order of $v$, following the
same steps presented in the previous sections, we obtain 
\begin{align}
E_{MC} & =\frac{\mu\sigma^{2}}{8\pi^{2}}\int d^{2}{\bf {p}}_{p}\left\{ 1+\frac{\left({\bf {p}}_{p}\cdot{\bf {v}}_{p}\right)^{2}}{2\left({\bf {p}}_{p}^{2}+m^{2}\right)}\left[\frac{\left(4\sqrt{{\bf {p}}_{p}^{2}+m^{2}}+\mu\right)}{\left(2\sqrt{{\bf {p}}_{p}^{2}+m^{2}}+\mu\right)}+2a\sqrt{{\bf {p}}_{p}^{2}+m^{2}}\right]\right.\nonumber \\
 & \left.+\frac{(v^{3})^{2}}{2}\left[\frac{\left(4\sqrt{{\bf {p}}_{p}^{2}+m^{2}}+\mu\right)}{\left(2\sqrt{{\bf {p}}_{p}^{2}+m^{2}}+\mu\right)}-2a\sqrt{{\bf {p}}_{p}^{2}+m^{2}}\right]\right\} \frac{e^{-2a\sqrt{{\bf {p}}_{p}^{2}+m^{2}}}}{2\sqrt{{\bf {p}}_{p}^{2}+m^{2}}\left(2\sqrt{{\bf {p}}_{p}^{2}+m^{2}}+\mu\right)}\ ,\label{intmc2}
\end{align}
where $a>0$ is the distance between the mirror and the charge. The
sub-index $MC$ means that we have the interaction energy between
the mirror and the charge.

Equation (\ref{intmc2}) can be simplified by using polar coordinates,
integrating out in the solid angle and performing the change of integration
variable $p\rightarrow y=2\sqrt{p^{2}+m^{2}}$ where $\mid{\bf {p}}_{p}\mid=p$,
\begin{align}
E_{MC} & =\frac{\mu\sigma^{2}}{16\pi}\int_{2m}^{\infty}dy\frac{e^{-ay}}{\left(y+\mu\right)}\Biggl[1+{\bf {v}}_{p}^{2}\frac{1}{y^{2}}\left(\frac{y^{2}}{4}-m^{2}\right)\left(\frac{\left(2y+\mu\right)}{\left(y+\mu\right)}+ay\right)\nonumber \\
 & +\frac{(v^{3})^{2}}{2}\left(\frac{\left(2y+\mu\right)}{\left(y+\mu\right)}-ay\right)\Biggr]\ .\label{intmc3}
\end{align}
The relevant integrals can be found in\,\cite{Gard}, 
\begin{equation}
\int_{2m}^{\infty}dy\frac{e^{-ay}}{\left(y+\mu\right)}=e^{\mu a}Ei\left(1,2ma+\mu a\right)\thinspace,\label{intmc5}
\end{equation}
\begin{align}
\int_{2m}^{\infty}dy\frac{e^{-ay}}{y^{2}\left(y+\mu\right)}\left(\frac{y^{2}}{4}-m^{2}\right)\left(\frac{\left(2y+\mu\right)}{\left(y+\mu\right)}+ay\right) & =\frac{1}{2}e^{\mu a}Ei\left(1,2ma+\mu a\right)\ ,
\end{align}
and 
\begin{align}
\int_{2m}^{\infty}dy\frac{e^{-ay}}{\left(y+\mu\right)}\left(\frac{\left(2y+\mu\right)}{\left(y+\mu\right)}-ay\right) & =2\Bigg[\left(\mu a+1\right)e^{\mu a}Ei\left(1,2ma+\mu a\right)\nonumber \\
 & \ -\frac{(m+\mu)}{(2m+\mu)}e^{-2ma}\Bigg]\ ,
\end{align}
where $Ei\left(u,s\right)$ is the exponential integral function\,\cite{Arfken}
defined by 
\begin{equation}
Ei(n,s)=\int_{1}^{\infty}\frac{e^{-ts}}{t^{n}}\ dt\ \ \,\ \ \ \Re(s)>0\ ,\ n=0,1,2,\cdots\ ,
\end{equation}
which can be extended by analytic continuation as follows 
\begin{equation}
Ei(n,s)=s^{n-1}\Gamma(1-n,s)\ ,
\end{equation}
$\Gamma\left(m,s\right)$ being the incomplete Gamma function.

As a result, the interaction energy reads 
\begin{align}
E_{MC}= & \frac{\mu\sigma^{2}}{16\pi}\Biggl\{ e^{\mu a}Ei\left(1,2ma+\mu a\right)+\frac{{\bf {v}}_{p}^{2}}{2}e^{\mu a}Ei\left(1,2ma+\mu a\right)\nonumber \\
 & +(v^{3})^{2}\left[\left(\mu a+1\right)e^{\mu a}Ei\left(1,2ma+\mu a\right)-\frac{(m+\mu)}{(2m+\mu)}e^{-2ma}\right]\Biggr\}\ .\label{energym1}
\end{align}
This is a perturbative result and gives the interaction energy between
a point-like scalar charge and a semi-transparent mirror in the massive
case. The first term on the right hand side reproduces the result
of the standard (Lorentz invariant) scalar field \cite{fabarone4},
the remaining terms are corrections due to the Lorentz symmetry breaking.

From the energy (\ref{energym1}), we derive two kinds of physical
phenomena. The first one is a force between the mirror and the charge,
\begin{align}
F_{MC}= & -\frac{\partial E_{MC}}{\partial a}=-\frac{\mu\sigma^{2}}{16\pi a}\Bigg[\left(1+\frac{{\bf {v}}_{p}^{2}}{2}\right)\Bigg(\mu ae^{\mu a}Ei\left(1,2ma+\mu a\right)-e^{-2ma}\Bigg)\label{FMC}\\
 & \ +(v^{3})^{2}\Bigg((2+\mu a)\mu ae^{\mu a}Ei\left(1,2ma+\mu a\right)-(\mu a+1)e^{-2ma}+2\frac{m+\mu}{2m+\mu}mae^{-2ma}\Biggr)\Bigg]\ ,
\end{align}
which is always attractive, provided that ${\bf {v}}_{p}^{2},(v^{3})^{2}<<1$.

\begin{figure}[!h]
\centering \includegraphics[scale=0.5]{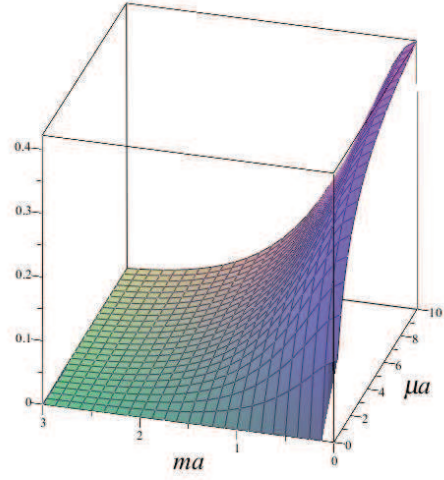} \caption{Function ${\cal F}_{p}$, appearing in the force described in Eq.\,\eqref{FMC2},
where the vertical axis is in arbitrary units.}
\label{graficocalF|}
\end{figure}

\begin{figure}[!h]
\centering \includegraphics[scale=0.5]{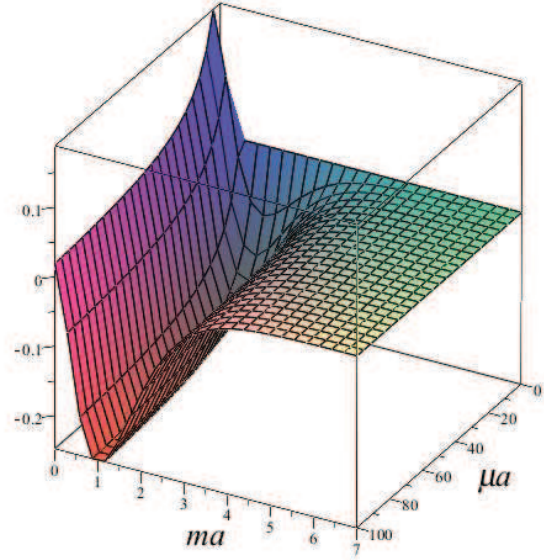} \caption{Function ${\cal F}_{3}$, appearing in the force described in Eq.\,\eqref{FMC2},
where the vertical axis is in arbitrary units.}
\label{graficocalFperp}
\end{figure}

Let us define the following dimensionless functions, 
\begin{align}
{\cal F}_{p}(x,y) & =\frac{x}{2}\left[e^{-2y}-xe^{x}Ei\left(1,2y+x\right)\right]\thinspace,\\
{\cal F}_{3}(x,y) & =x\left[(x+1)e^{-2y}-\left(x+2\right)xe^{x}Ei\left(1,y+x\right)-2\frac{(y+x)}{(2y+x)}ye^{-2y}\right]\thinspace,
\end{align}
and rewrite the force (\ref{FMC}) in the form 
\begin{equation}
F_{MC}=\frac{\sigma^{2}}{16\pi}\frac{1}{a^{2}}\Bigg[\mu a\Big(e^{-2ma}-\mu ae^{\mu a}Ei\left(1,2ma+\mu a\right)\Big)+\frac{{\bf {v}}_{p}^{2}}{2}{\cal F}_{p}(\mu a,ma)+(v^{3})^{2}{\cal F}_{3}(\mu a,ma)\Bigg]\thinspace,\label{FMC2}
\end{equation}
where we have a Coulombian behavior modulated by the expression inside
brackets. The correction due to the Lorentz symmetry breaking is given
by the functions ${\cal F}_{p}$ and ${\cal F}_{3}$, the first one
is associated with the components of the background vector parallel
to the mirror and the second one, with the component perpendicular
to the mirror. ${\cal F}_{p}$ is positive in its domain and ${\cal F}_{3}$
assume positive and negative values, as shown in Fig.\,\ref{graficocalF|}
and\,\ref{graficocalFperp}. Both functions vanish in the limit $\mu=0$,
where we have no mirror present.

\begin{figure}[!h]
\centering \includegraphics[scale=0.5]{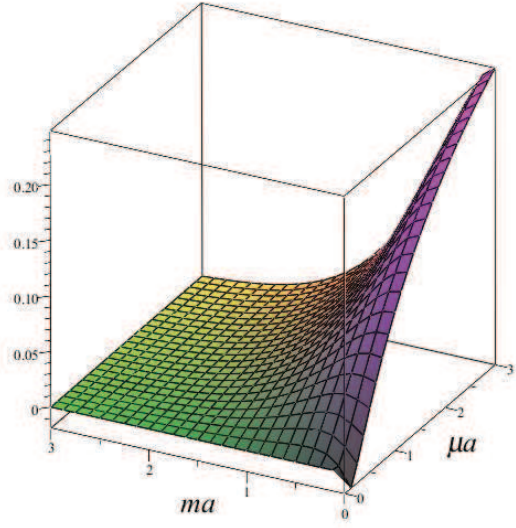} \caption{Function ${\cal T}$ of Eq. (\ref{defcalT}), where the vertical axis
is in arbitrary units.}
\label{calT}
\end{figure}

\begin{figure}[!h]
\centering \includegraphics[scale=0.4]{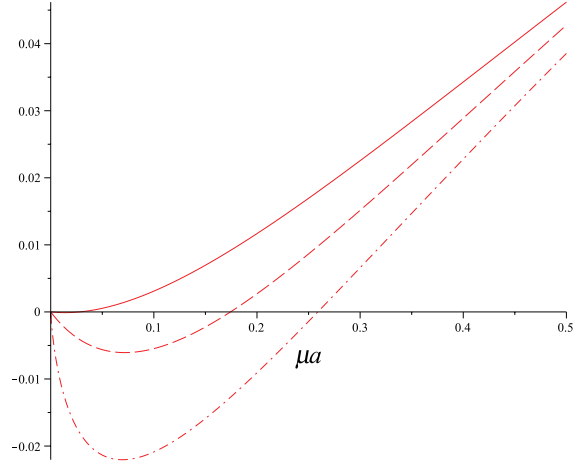} \caption{Graph of the function ${\cal T}$ for $ma=0$ (dash-point), $ma=0.1$
(dash) and $ma=0.2$ (solid) as a function of $\mu a$, where the
vertical axis is in arbitrary units.}
\label{calT2}
\end{figure}

The second phenomena is obtained when we fix the distance between
the charge and the mirror and vary the orientation of the whole system
with respect to the background vector. In this case, we can show that
a torque emerges on the system. In order to calculate this torque,
we define as $0\leq\alpha\leq\pi$ the angle between the normal to
the mirror ($\hat{x}^{3}$) and the background vector, in such a way
that 
\begin{equation}
(v^{3})^{2}={\bf v}^{2}\cos^{2}(\alpha)\ \ ,\ \ {\bf {v}}_{p}^{2}={\bf v}^{2}\sin^{2}(\alpha)\ ,
\end{equation}
then the torque can be computed from Eq.\,(\ref{energym1}) as follows,
\begin{equation}
\tau_{MC}=-\frac{\partial E_{MC}}{\partial\alpha}=-\frac{\mu\sigma^{2}{\bf v}^{2}}{16\pi}\sin(2\alpha)\left[\left(\mu a+\frac{1}{2}\right)e^{\mu a}Ei\left(1,2ma+\mu a\right)-\frac{(m+\mu)}{(2m+\mu)}e^{-2ma}\right]\thinspace.\label{torqueMC}
\end{equation}
Equation (\ref{torqueMC}) is a new effect, which disappears in the
$v=0$ limit. Defining the function 
\begin{equation}
{\cal T}(x,y)=x\left[\frac{(y+x)}{(2y+x)}e^{-2y}-\left(x+\frac{1}{2}\right)e^{x}Ei\left(1,2y+x\right)\right]\thinspace,\label{defcalT}
\end{equation}
we can write Eq.\,(\ref{torqueMC}) in the form 
\begin{equation}
\tau_{MC}=\frac{\sigma^{2}{\bf v}^{2}}{16\pi}\frac{1}{a}\sin(2\alpha){\cal T}(\mu a,ma)\ .
\end{equation}
In Fig.\,(\ref{calT}), we show the behavior of ${\cal T}$ in terms
of $\mu a$ and $ma$. The function is positive except in a very small
region around $\mu a=ma=0$, and goes to zero if $ma$ is large or
$\mu a$ approaches zero. This behavior can also be seen in Fig.\,(\ref{calT2}),
where we have three plots, with three different values for the mass,
in the vicinity of $\mu a=0$. In the limit $\mu a\to0$, the result
in Eq.\,(\ref{defcalT}) vanishes, as expected. This torque and the
force modulation contained in Eq.\,\eqref{FMC2} are phenomenological
signatures of the Lorentz violation introduced by the $v^{\mu}$,
and might be relevant in experimental setups involving mirrors.

For the massless case, the energy (\ref{energym1}) becomes 
\begin{equation}
E_{MC}\left(m=0\right)=\frac{\mu\sigma^{2}}{16\pi}\Biggl\{ e^{\mu a}Ei\left(1,\mu a\right)+\frac{{\bf {v}}_{p}^{2}}{2}e^{\mu a}Ei\left(1,\mu a\right)+(v^{3})^{2}\left[\left(\mu a+1\right)e^{\mu a}Ei\left(1,\mu a\right)-1\right]\Biggr\}\ .\label{energy1m0}
\end{equation}

The limit $\mu\rightarrow\infty$ is interesting, corresponding physically
to the field subjected to Dirichlet boundary conditions in the plane.
In this limit, we have a perfect two-dimensional mirror and, from
Eq. (\ref{energym1}), we obtain 
\begin{equation}
E_{MC}\left(\mu\rightarrow\infty\right)=\frac{\sigma^{2}}{16\pi}\frac{e^{-2ma}}{a}\left(1+\frac{{\bf {v}}_{p}^{2}}{2}-ma(v^{3})^{2}\right)\ .\label{energym3}
\end{equation}
The first term on the right hand side is the three-dimensional Yukawa
potential between two charges at a distance $2a$ apart. The second
and third terms are corrections due to the Lorentz symmetry breaking
up to second order in $v^{\mu}$. The corresponding interaction force
between the point-like charge and the perfect mirror is given by 
\begin{equation}
F_{MC}\left(\mu\rightarrow\infty\right)=-\frac{\partial E_{MC}\left(\mu\rightarrow\infty\right)}{\partial a}=\frac{\sigma^{2}}{8\pi}\frac{e^{-2ma}}{a}\left[\left(1+\frac{{\bf {v}}_{p}^{2}}{2}\right)\left(m+\frac{1}{2a}\right)-m^{2}a(v^{3})^{2}\right]\thinspace.\label{forcem4}
\end{equation}

In Eq. (\ref{force1}) we have the interaction force between two point-like
scalar charges for the model (\ref{model1}). Expanding this expression
up to second order in $v^{\mu}$, we can obtain the interaction force
for the special case where we have two opposite point-like charges,
$\sigma_{1}=\sigma$ and $\sigma_{2}=-\sigma$, placed at a distance
$2a$ apart. In this specific situation, this force turns out to be
equivalent to Eq. (\ref{forcem4}). The interesting conclusion is
that the image method is valid for the Lorentz violation theory (\ref{model1})
up to second order in $v^{\mu}$ for the Dirichlet boundary condition.

Taking the limit when $\mu\rightarrow\infty$ in Eq. (\ref{energy1m0})
or equivalently putting $m=0$ in (\ref{energym3}), we obtain the
interaction energy between a point charge and a perfect mirror for
the massless scalar field, and consequently the interaction force,
\begin{equation}
F_{MC}\left(\mu\rightarrow\infty,m=0\right)=\frac{\sigma^{2}}{16\pi a^{2}}\left(1+\frac{{\bf {v}}_{p}^{2}}{2}\right)\ ,\label{forcem5}
\end{equation}
which is the usual Coulombian force with an overall minus sign between
the scalar charge and its image, placed at a distance $2a$ apart.
With the same analysis, one can argue that Eq. (\ref{forcem5}) is
in agreement with Eq. (\ref{for1exp}), and again the validity of
the image method is verified. In the same limit, from Eq.\,(\ref{torqueMC}),
we have 
\begin{equation}
\tau_{MC}\left(\mu\rightarrow\infty,m=0\right)=-\frac{\partial E_{MC}\left(\mu\rightarrow\infty,m=0\right)}{\partial\alpha}=\frac{\sigma^{2}{\bf {v}}^{2}}{32\pi a}\sin(2\alpha)\ .\label{torqueMC2}
\end{equation}
When $\alpha=0,\pi/2,\pi$, corresponding to the mirror being parallel,
perpendicular or antiparallel to the background vector ${\bf v}$,
the torque (\ref{torqueMC2}) vanishes. The configurations $\alpha=0,\pi$
are stable equilibrium situations, while for $\alpha=\pi/2$ we have
an unstable equilibrium point. When $\alpha=\pi/4,3\pi/4$, the torque
(\ref{torqueMC2}) exhibits its maximum and minimum values, respectively.
The equilibrium situation is attained when the mirror is parallel
or antiparallel to the background vector.

\subsection{Exact results\label{Interaction3}}

The first case in which we can provide exact results is when $v^{\mu}=v_{p}^{\mu}$,
what leads to 
\begin{equation}
E_{MC}=\frac{\mu\sigma^{2}}{8\pi^{2}}\int d^{2}{\bf {p}}_{p}\frac{e^{-2a\sqrt{{\bf {p}}_{p}^{2}-\left({\bf {p}}_{p}\cdot{\bf {v}}_{p}\right)^{2}+m^{2}}}}{2\sqrt{{\bf {p}}_{p}^{2}-\left({\bf {p}}_{p}\cdot{\bf {v}}_{p}\right)^{2}+m^{2}}\left(2\sqrt{{\bf {p}}_{p}^{2}-\left({\bf {p}}_{p}\cdot{\bf {v}}_{p}\right)^{2}+m^{2}}+\mu\right)}\ .\label{energyem1}
\end{equation}
Performing a change in the integration variables similar to the one
we have made in the Appendix, and then using polar coordinates, we
have 
\begin{equation}
E_{MC}=\frac{\mu\sigma^{2}}{4\pi\sqrt{1-{\bf {v}}_{p}^{2}}}\int_{0}^{\infty}dq\ q\frac{e^{-2a\sqrt{q^{2}+m^{2}}}}{2\sqrt{q^{2}+m^{2}}\left(2\sqrt{q^{2}+m^{2}}+\mu\right)}\ .\label{energyem2}
\end{equation}
Now, carrying out the change of integration variable $y=2\sqrt{q^{2}+m^{2}}$,
we obtain 
\begin{equation}
E_{MC}=\frac{\mu\sigma^{2}}{16\pi\sqrt{1-{\bf {v}}_{p}^{2}}}e^{\mu a}Ei\left(1,2ma+\mu a\right)\ .\label{energyem3}
\end{equation}
Equation (\ref{energyem3}) gives the exact expression for the interaction
energy between a point-like current and a semi-transparent mirror
for the special case where the background vector has only the parallel
components to the mirror. We notice that (\ref{energyem3}) is the
usual result found in standard scalar field theory with an effective
coupling constant $\sigma\rightarrow\sigma\left(1-{\bf {v}}_{p}^{2}\right)^{-1/4}$.
Taking the limit $\mu\rightarrow\infty$ in Eq. (\ref{energyem3})
and computing the interaction force, we arrive at 
\begin{eqnarray}
F_{MC}\left(\mu\rightarrow\infty\right)=\frac{\sigma^{2}}{16\pi\sqrt{1-{\bf {v}}_{p}^{2}}}\frac{e^{-2ma}}{a}\left(2m+\frac{1}{a}\right)\ ,\label{forceem1}
\end{eqnarray}
which is the interaction force characterized by the Dirichlet's boundary
condition.

In Eq. (\ref{force1}) we have the exact interaction force between
two point-like currents. For the special situation where $v^{3}=0,\sigma_{1}=\sigma,\sigma_{2}=-\sigma$
and $a\to2a$, this result turns out to be equivalent to Eq. (\ref{forceem1}).
Thus, we again verify that for this special case, $(v^{3}=0)$, the
image method is valid.

The second exact case we discuss is when only $v^{3}$ is nonzero,
what leads to the result 
\begin{equation}
E_{MC}=\frac{\sigma^{2}}{16\pi}\frac{\mu}{\left[1-(v^{3})^{2}\right]}e^{\mu a\left[1-(v^{3})^{2}\right]^{-1}}Ei\left(1,2ma\left[1-(v^{3})^{2}\right]^{-1}+\mu a\left[1-(v^{3})^{2}\right]^{-1}\right)\ .\label{energyev31}
\end{equation}
Eq. (\ref{energyev31}) is equivalent to the result obtained in standard
scalar field theory with an effective mass $m\rightarrow m\left[1-(v^{3})^{2}\right]^{-1}$
and an effective degree of transparency of the mirror $\mu\rightarrow\mu\left[1-(v^{3})^{2}\right]^{-1}$.
From Eq. (\ref{energyev31}) we can compute the interaction force
in the limit $\mu\rightarrow\infty$, resulting in 
\begin{eqnarray}
F_{MC}\left(\mu\rightarrow\infty\right)=\frac{\sigma^{2}}{16\pi}\frac{e^{-2m\left[1-(v^{3})^{2}\right]^{-1}a}}{a}\left(\frac{1}{a}+2m\left[1-(v^{3})^{2}\right]^{-1}\right)\ .\label{forcev32}
\end{eqnarray}
For the massless case, the interaction force (\ref{forcev32}) becomes
the corresponding Coulombian interaction between two charges at a
distance $2a$ apart with an overall minus sign. Thus, in this particular
scenario, Lorentz violation effects disappear from the end result.
As before, taking ${\bf {v}}_{p}^{\mu}=0,\sigma_{1}=\sigma,\sigma_{2}=-\sigma$
and $a\to2a$ in Eq.\,(\ref{force1}), we reproduce the result in
Eq.\,(\ref{forcev32}). Thus, the image method is also valid for
the case where $v^{\mu}=\left(0,0,0,v^{3}\right)$.

It is important to mention that the validity of the image method in
a Lorentz-violating scenario is a non-trivial result, since the presence
of the LV background reduces the symmetry of the problem, which is
a key element in the application of the method. This suggests that
the presence of mirrors in Lorentz-violating scenarios is a subject
which deserves more investigation.

\section{Final Remarks\label{conclusions}}

In this paper, we investigated the interactions between external sources
for a massive real scalar field in the presence of an aether-like
CPT-even Lorentz symmetry breaking term. First we performed an analysis
in $\dimp+D+1$ dimensions where we considered steady field sources
concentrated along parallel $D$-branes, without recourse to any approximation
schemes. We discussed some particular instances of our general results
and observed effects with no counterpart in the standard (Lorentz
invariant) scalar field theory. For example, we have shown the emergence
of an spontaneous torque on a classical scalar dipole which is an
exclusive effect due to the Lorentz symmetry breaking, agreeing with
results obtained in different, more complicated models such as\,\cite{borges1}.

Afterwards, some consequences of the Lorentz violation theory (\ref{model1})
due to the presence of a semi-transparent mirror were studied in $3+1$
dimensions. We considered different configurations of the background
vector, starting by taking into account all the components of the
background vector, and treating it perturbatively up to second order.
Next, we provided exact results for two special cases, specifically
when the background vector has only components parallel and perpendicular
to the mirror. For all these configurations of the background vector,
we obtained the propagator for the scalar field and the interaction
force between the mirror and a point-like current. We showed that
the image method is valid in the considered theory for Dirichlet boundary
condition. We also showed that a new effect arises from the obtained
results, a torque acting on the mirror according to its positioning
with respect to the background vector.

These results suggest that the extension of these studies to more
general LV models is a very interesting prospect. Despite not being
directly applicable to the phenomenological search of Lorentz violation
established within the formalism of the Standard Model extension\cite{SME1,SME2,SME3,SME4},
the scalar field can still be explored as a prototype, establishing
interesting effects of LV yet to be explored. A first natural extension
of our results would be to more general LV backgrounds as described
by Eq.\,\eqref{genLVmin}. The extension of these studies for non-minimal
(higher-derivative) LV models would also be of interest.

\textbf{\medskip{}
}

\textbf{Acknowledgments.} The authors would like to thank V. A. Kostelecky
for reading the manuscript and providing important feedback. This
work was partially supported by Conselho Nacional de Desenvolvimento
Científico e Tecnológico (CNPq) and Fundação de Amparo à Pesquisa
do Estado de São Paulo (FAPESP), via the following grants: CNPq 311514/2015-4,
CNPq 313978/2018-2 (FAB), CNPq 304134/2017-1 and FAPESP 2017/13767-9
(AFF), FAPESP 2016/11137-5 (L.H.C.B.).

\appendix

\section{The Eqs. (\ref{exactG1}) and (\ref{exactG1v3}) \label{APPENDIX}}

In this appendix we provide additional details on the computation
of Eqs. (\ref{exactG1}) and (\ref{exactG1v3}). We note that in some
of the intermediate expressions that follow, the condition $v^{2}=0$
cannot be imposed to ensure the tracelessness of the LV coefficient
$k^{\mu\nu}$ defined in Eq.\,\eqref{genLVmin}; however, this condition
can be safely imposed in the final result, from which one can obtain,
in the proper limiting cases, the perturbative results previously
obtained, thus ensuring the consistency of the calculation.

Starting from Eq.\,(\ref{propagator}), in order to put $G_{0}\left(x,y\right)$
in an appropriated form, we have to carry out a change of the integration
variables similar to the ones employed in references \cite{LHCFAB1,scalar3}.
We split the four-vector momentum $p^{\mu}$ into two parts, one parallel,
${p}_{pa}^{\mu}$, and the other normal, ${p}_{no}^{\mu}$, to the
Lorentz violation parameter $v^{\mu}$, 
\begin{equation}
{p}^{\mu}={p}_{no}^{\mu}+{p}_{pa}^{\mu}\ ,\ {p}=\Bigl(\frac{{v}\cdot{p}}{{v}^{2}}\Bigr){v^{\mu}}\ ,\ {p}_{no}^{\mu}={p}^{\mu}-\Bigl(\frac{{v}\cdot{p}}{{v}^{2}}\Bigr){v^{\mu}}\ ,\label{mudan1EM22}
\end{equation}
where ${p}_{no}\cdot{v}=0$ and $\left(p\cdot v\right)^{2}=p_{pa}^{2}v^{2}$
. Now, we define the four-vector ${q^{\mu}}$ 
\begin{equation}
{q}^{\mu}={p}_{no}^{\mu}+{p}_{pa}^{\mu}\sqrt{1+{v}^{2}}={p}^{\mu}+\Bigl(\frac{{v}\cdot{p}}{{v}^{2}}\Bigr)(\sqrt{1+{v}^{2}}-1)v^{\mu}\ .\label{defq33}
\end{equation}
With definitions (\ref{mudan1EM22}) and (\ref{defq33}), we have
\begin{equation}
{p}_{pa}^{\mu}=\frac{(v\cdot q)}{{v}^{2}}\frac{v^{\mu}}{\sqrt{1+{v}^{2}}}\ \ \ ,\ \ \ {p}_{no}^{\mu}=q^{\mu}-\frac{(v\cdot q)}{{v}^{2}}v^{\mu}\thinspace,\label{mudan6EM66}
\end{equation}
\[
{p}^{\mu}={q}^{\mu}+\frac{(v\cdot q)}{{v}^{2}}\left(\frac{1}{\sqrt{1+{v}^{2}}}-1\right)v^{\mu}\ ,
\]
and 
\begin{equation}
{q}^{2}={p}^{2}+({p}\cdot{v})^{2}\thinspace.\label{zxc222}
\end{equation}
With the aid of the definition 
\begin{equation}
{b}^{\mu}=\left({x^{\mu}-y^{\mu}}\right)+\left(\frac{1-\sqrt{1+{v}^{2}}}{\sqrt{1+{v}^{2}}}\right)\left(\frac{{v}\cdot\left({x-y}\right)}{{v}^{2}}\right)v^{\mu}\ ,\label{zxc355}
\end{equation}
and Eq. (\ref{mudan6EM66}), we obtain 
\begin{equation}
{p}\cdot{\left(x-y\right)}={b}\cdot{q}\label{mudan3EM44}
\end{equation}
The Jacobian of the transformation from ${p}^{\mu}$ to ${q}^{\mu}$
can be calculated from Eq. (\ref{mudan6EM66}) 
\begin{equation}
\det\left[\frac{\partial{p}^{\mu}}{\partial{q}_{\nu}}\right]=-\frac{1}{\sqrt{1+{v}^{2}}}\ .\label{mudan5EM88}
\end{equation}

Using these results, we obtain 
\begin{align}
G_{0}\left(x,y\right) & =-\frac{1}{\sqrt{1+{v}^{2}}}\int\frac{d^{4}q}{\left(2\pi\right)^{4}}\frac{e^{ib\cdot q}}{(q^{2}-m^{2})}\nonumber \\
 & =-\frac{1}{\sqrt{1+{v}^{2}}}\int\frac{d^{3}q_{p}}{\left(2\pi\right)^{3}}\ e^{ib_{p}\cdot q_{p}}\int\frac{dq^{3}}{2\pi}\frac{e^{-iq^{3}b^{3}}}{(q^{2}-m^{2})}\ .\label{intt40}
\end{align}
The first integral in Eq. (\ref{intt40}) is given by 
\begin{equation}
\int\frac{d^{3}q_{p}}{\left(2\pi\right)^{3}}\ e^{ib_{p}\cdot q_{p}}=-\sqrt{1+v_{p}^{2}}\int\frac{d^{3}p_{p}}{\left(2\pi\right)^{3}}\ e^{ip_{p}\cdot\left(x_{p}-y_{p}\right)}\ ,\label{intpar2}
\end{equation}
where we used the Eqs. (\ref{mudan3EM44}) and (\ref{mudan5EM88}),
while the last integral is given by 
\begin{equation}
\int\frac{dq^{3}}{2\pi}\frac{e^{-iq^{3}b^{3}}}{(q^{2}-m^{2})}=-\frac{e^{-L\mid b^{3}\mid}}{2L}\ ,\label{int50}
\end{equation}
where $L=\sqrt{m^{2}-q_{p}^{2}}$ or, from Eq. (\ref{zxc222}), $L=\sqrt{m^{2}-\left[p_{p}^{2}+\left(p_{p}\cdot v_{p}\right)^{2}\right]}$,
and $b^{3}$ is found by taking $\mu=3$ in (\ref{zxc355}), as follows,
\begin{equation}
b^{3}=\left(x^{3}-y^{3}\right)+\left(\frac{1-\sqrt{1+{v}^{2}}}{\sqrt{1+{v}^{2}}}\right)\left(\frac{{v}\cdot\left({x-y}\right)}{{v}^{2}}\right)v^{3}\ ,\label{b311}
\end{equation}

Collecting terms, we write 
\begin{equation}
G_{0}\left(x,y\right)=\int\frac{d^{3}p_{p}}{\left(2\pi\right)^{3}}\ e^{ip_{p}\cdot\left(x_{p}-y_{p}\right)}\left[-\frac{1}{2}\ \sqrt{\frac{1+v_{p}^{2}}{1+v^{2}}}\frac{e^{-L\mid b^{3}\mid}}{L}\right]\ .\label{inttt}
\end{equation}
Finally, taking $v^{3}=0$ in the term between brackets on the right-hand
side of the Eq. (\ref{inttt}), we obtain the Eq. (\ref{exactG1}).
In the same way, taking $v_{p}^{\mu}=0$ , we obtain Eq. (\ref{exactG1v3}).

\end{document}